\documentclass[runningheads]{llncs}
\usepackage[T1]{fontenc}
\usepackage{graphicx}
\usepackage{amsmath}
\usepackage{amssymb}

\usepackage{orcidlink}
\usepackage{tikz}
\usetikzlibrary{positioning, arrows.meta, shapes.geometric}

\definecolor{cnncolor}{RGB}{70,130,180}       % muted blue
\definecolor{vitcolor}{RGB}{60,150,90}        % muted green
\definecolor{aggcolor}{RGB}{120,90,160}       % muted purple
\definecolor{inputcolor}{RGB}{80,80,80}       % dark gray

\usepackage{tikz}
\usetikzlibrary{positioning, arrows.meta, shapes.geometric}
\usepackage{amsmath}
\usepackage{amssymb}

\definecolor{cnncolor}{RGB}{60,110,170}
\definecolor{vitcolor}{RGB}{50,140,80}
\definecolor{aggcolor}{RGB}{120,80,150}
\definecolor{mathcolor}{RGB}{30,30,30}

\begin{document}
\title{A Comparative and Hybrid Study of CNN and Transformer Models for Multi-Class Virus Classification in Transmission Electron Microscopy}
\titlerunning{CNN and Transformer Models for Multi-Class Virus Classification in TEM}
\author{Md Mahmudul Hoque$^*$\inst{1}\orcidlink{0000-0002-2618-4157},
Md Kawser Islam\inst{2}\orcidlink{0009-0001-2441-6815},
Md. Mamunur Rahman Moon\inst{3}\orcidlink{0009-0008-9557-0749},
Abdullah Rakib Akand\inst{4}\orcidlink{0009-0000-3708-6387},
Md. Hadi Al-amin\inst{5}\orcidlink{0009-0002-3334-8671} \and
H.M. Azrof\inst{6}\orcidlink{0009-0002-0287-5823}}
\authorrunning{Hoque et al.}
\institute{Dept. of Computer Science \& Engineering, CCN University of Science and Technology, CCN Road, Cumilla - 3506, Bangladesh.\\
\email{cse.mahmud.evan@gmail.com$^*$}\\ \and
Deakin University,221 Burwood Hwy, Burwood Vic 3125, Australia. \and
American International University-Bangladesh, Khilkhet, Dhaka 1229, Bangladesh.\\ \and
Dept. of CSE, Asian University of Bangladesh, Ashulia, Dhaka 1341, Bangladesh. \\ \and
Department of Mechatronics, Magura polytechnic Institute , Magura, khulna Bangladesh 7600. \\ \and
Department of Computer Science and Engineering,
Khulna University of Engineering \& Technology (KUET),Khulna, Bangladesh.}

\maketitle              % typeset the header of the contribution
\begin{abstract}
The automatic recognition of virus particles in transmission electron microscopy (TEM) images remains a demanding task, primarily owing to strong inter-class similarity, scale variability, and pronounced class imbalance. In this study, several convolutional neural networks and transformer-based architectures were comparatively evaluated for the classification of 22 virus categories using the TEM virus dataset. All models were trained under identical preprocessing and optimization conditions, and imbalance effects were mitigated through a weighted cross-entropy formulation. Performance was quantified using overall accuracy together with macro-averaged precision, recall, and F1 score. Among standalone models, the Swin Transformer achieved the highest accuracy (0.8831) and macro-F1 score (0.8444), followed by DeiT (accuracy 0.8669). Convolutional architectures exhibited comparatively lower balanced performance, with ResNet50 demonstrating substantial degradation (accuracy 0.5887) under imbalanced conditions. To exploit complementary representational properties, decision-level hybrid strategies were implemented. The performance-weighted hybrid attained an accuracy of 0.8831 and the highest macro-F1 score (0.8528), slightly surpassing the equal-weight hybrid configuration. These observations indicate that architectural heterogeneity contributes to improved inter-class balance without sacrificing overall predictive accuracy. Future work may explore scale-aware representations, feature-level fusion mechanisms, and expanded TEM datasets to further enhance robustness and generalization in virus identification tasks.

\end{abstract}
\section{Introduction}
Viruses consist of short strands of nucleic acid—either DNA or RNA—encapsulated within a protective protein shell. Owing to their nanoscale dimensions, direct visualization was not feasible prior to the development of electron microscopy technologies \cite{taylor2014virus}. Among available imaging modalities, transmission electron microscopy (TEM) remains uniquely capable of directly resolving viral particles, as its nanometer-scale resolution exceeds the diffraction limits inherent to light microscopy \cite{roingeard2019virus}. By achieving magnifications of several million times, TEM enables detailed visualization of viral morphology, including structural features that are critical for taxonomic differentiation and diagnostic assessment.

In parallel with advances in imaging technology, deep learning methodologies have become increasingly prominent in medical image classification. Convolutional neural networks (CNNs), in particular, have demonstrated strong performance in virological and biomedical imaging tasks \cite{hoque2026computervisionhybridapproach,hu2025flu,ma2021multi,wang2020differentiation,yin2021iav}. For example, two-dimensional CNN architectures have been employed for predicting antigenic variants and clustering seasonal influenza A viruses \cite{yin2021iav,meng2024predac,amin2024sentiment}. The introduction of distributed amino acid representations such as ProtVec further enabled proteomic feature learning, while the PREDAC-CNN framework was designed to model antigenic evolution patterns, with reported robustness in variant prediction. More recently, transformer-based architectures have been introduced into medical imaging and viral classification contexts \cite{shiraj2024study,shah2024deng,hoque2026visionmodelsmedicalimaging}. Vision Transformers have been utilized for tasks ranging from UAV-based mapping of plant viral infections to lineage assignment of SARS-CoV-2 genomes, where models such as ViRAL achieved top-1 accuracies exceeding 94\% under optimized deployment conditions \cite{jamali2025high,jahshan2024viral,hoque2023analyzing}. Hybrid approaches integrating CNNs and transformers have also been explored in medical imaging and disease detection domains \cite{shafi2024hybrid,cao2025efficientmfrf,omololu2hybrid,shao2022novel,naidji2024novel,maheshwari2025automated,iqbal2025rs}. Such architectures aim to combine the strong local feature extraction capacity of CNNs with the global contextual modeling capabilities of self-attention mechanisms. Despite these advances, relatively limited research has focused specifically on TEM-based virus image analysis using deep learning techniques. Early investigations addressed structural characterization and critical imaging considerations \cite{zaefferer2011critical,hoque2024comparison}, while more recent studies have begun to apply CNN-based and related models to the TEM virus dataset \cite{matuszewski2021temm,singh2025virus,chen2024advancing}. Nevertheless, substantial challenges remain, particularly in addressing inter-class morphological similarity, scale variability, and pronounced class imbalance. Consequently, further methodological refinement and comprehensive comparative evaluation remain necessary for robust multi-class virus identification in TEM imagery.

\section{Methodology}
\begin{figure}
    \centering
    \includegraphics[width=0.5\linewidth]{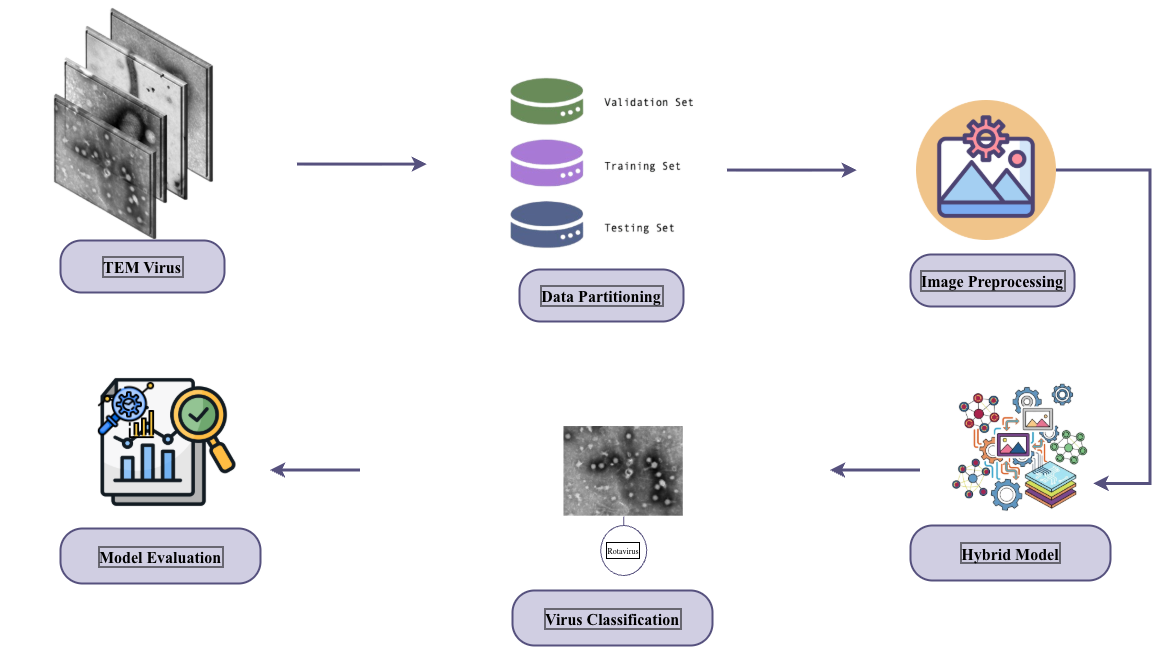}
    \caption{System Design}
    \label{fig.system}
\end{figure}

The Fig. \ref{fig.system} shows the proposed system. The proposed system begins with TEM Virus images undergoing Image Preprocessing to enhance quality, followed by Data Partitioning into training and testing sets. A Hybrid Model is then employed for Virus Classification, with its performance rigorously assessed through Model Evaluation to ensure accurate and reliable results.

\subsection{Data Collection}
\begin{figure}
    \centering
    \includegraphics[width=0.31\linewidth]{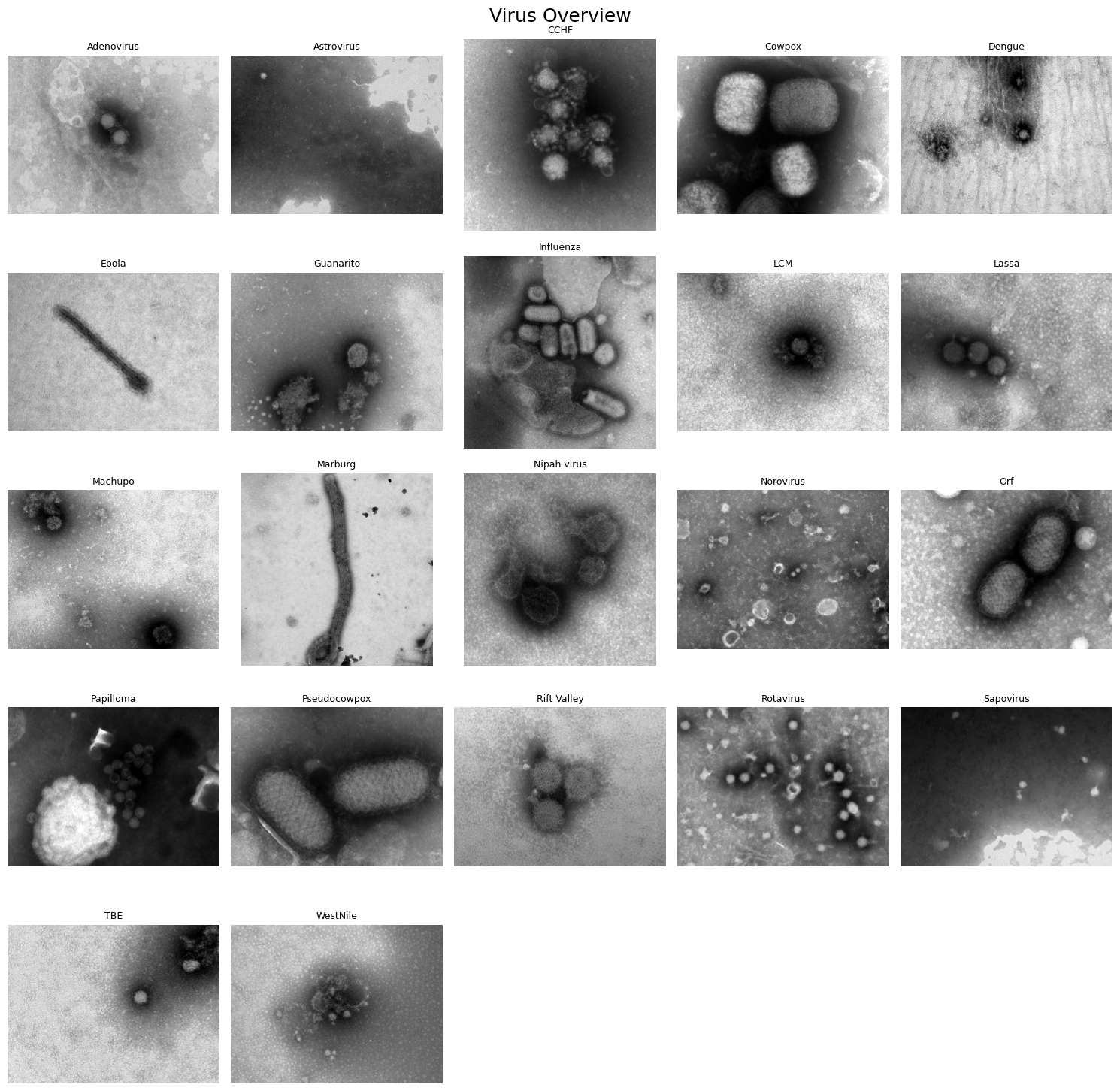}
    \caption{Virus Overview}
    \label{fig:virus}
\end{figure}

The TEM virus dataset \cite{matuszewski2021tem} comprises 1,245 transmission electron microscopy images spanning 22 virus classes. Image acquisition was carried out using two electron microscopy systems, namely a LEO microscope (Zeiss) equipped with a Morada camera (Olympus) and a Tecnai 10 system (FEI) coupled with a MegaView III camera (Olympus). A pronounced class imbalance is present, with image counts ranging from 9 to 129 per class and annotated particle instances varying between 38 and 1,934. The dataset exhibits notable heterogeneity in spatial characteristics. Image resolutions are either 1376 × 1032 or 2048 × 2048 pixels, and pixel sizes range from 0.26 nm to 5.57 nm, reflecting differences in acquisition magnification. Annotation data are stored in separate text files corresponding to each image. Representative examples of all virus classes are illustrated in Fig.~\ref{fig:virus}, highlighting the considerable morphological diversity and structural complexity inherent in TEM imagery.

\subsection{Data Preprocessing}
Owing to variability in spatial resolution, magnification, and class distribution, a preprocessing pipeline was applied to improve consistency while preserving morphological characteristics of viral structures.
\subsubsection{Image Standardization and Resizing}
The raw images, originally acquired at resolutions of 1376 × 1032 and 2048 × 2048 pixels, were resized to 224 × 224 pixels to ensure compatibility with CNN and transformer architectures. Bilinear interpolation was employed to maintain structural continuity. Although pixel sizes ranged from 0.26 nm to 5.57 nm, no explicit scale normalization was performed, allowing models to implicitly learn scale-invariant representations.
\subsubsection{Channel Adaptation and Normalization}
Since TEM images are grayscale, each image was converted to a three-channel format via channel replication to match pretrained model requirements. Pixel values were scaled to $[0,1]$, followed by normalization using ImageNet mean and standard deviation, thereby aligning input distributions with pretrained weights.
\subsubsection{Data Augmentation}
To mitigate overfitting and address limited samples in certain classes, data augmentation was applied during training. Transformations included random horizontal and vertical flipping, mild rotations, and random resized cropping. These operations introduced controlled variability without distorting essential morphological features.
\subsubsection{Handling Class Imbalance}
Significant class imbalance was addressed using a weighted cross-entropy loss. Class weights were computed via inverse-frequency scaling,
\begin{equation}
w_c = \frac{N}{n_c},
\end{equation}
and normalized as
\begin{equation}
\tilde{w}_c = \frac{w_c}{\sum_{k=1}^{C} w_k} \cdot C.
\end{equation}
The loss function was defined as
\begin{equation}
\mathcal{L} = - \sum_{c=1}^{C} \tilde{w}_c \, y_c \log(\hat{y}_c),
\end{equation}
where $y_c$ and $\hat{y}_c$ denote the ground-truth and predicted probabilities, respectively. This formulation increases the contribution of underrepresented classes during optimization.

\section{Model Training and Development}

\subsection{Network Architectures}
To investigate heterogeneous representation learning, both convolutional and transformer-based architectures were considered. ResNet50, EfficientNetV2, and ConvNeXt were employed as CNN backbones, while Vision Transformer (ViT), Swin Transformer, and DeiT represented transformer-based models. For all architectures, the final classification layers were replaced with linear layers corresponding to the 22 virus classes. Backbone parameters were initialized with ImageNet-pretrained weights where available, while classification layers were trained from scratch.

\subsection{Training Protocol}
All models were trained under identical experimental settings to ensure comparability. A stratified split was applied to preserve class distribution across training and validation sets. Optimization was performed using AdamW with an initial learning rate of $1 \times 10^{-4}$, followed by cosine annealing for gradual learning rate decay. Parameter updates were carried out using mini-batch gradient descent, with weighted cross-entropy loss employed to address class imbalance.

\subsection{Regularization and Optimization Strategy}
To mitigate overfitting arising from limited samples in several classes, data augmentation was applied during training, introducing moderate spatial variability. Weight decay was incorporated to constrain parameter magnitudes and improve generalization. Model selection was based on early stopping using validation macro-F1 score, which was preferred over accuracy due to class imbalance. The checkpoint achieving the highest F1 score was retained for subsequent evaluation.

\subsection{Hybrid CNN--Transformer Architecture}
To exploit complementary model behavior, decision-level aggregation was implemented. Posterior probabilities from individual models were combined using both equal-weight and performance-weighted averaging. Final predictions were obtained by selecting the class with the highest aggregated probability. Rather than performing intermediate feature fusion, a probabilistic decision-level aggregation framework was adopted. Let $\mathcal{M} = \{M_1, M_2, \dots, M_K\}$ denote the collection of trained models, where $K=6$ in the present study. For an input image $x$, each model produces a logit vector $z_k(x) \in \mathbb{R}^{C}$, where $C=22$.

Class posterior probabilities are computed using the softmax transformation:
\begin{equation}
p_k(c \mid x) = \frac{\exp(z_{k,c}(x))}{\sum_{j=1}^{C} \exp(z_{k,j}(x))}.
\end{equation}

Under equal-weight aggregation, the hybrid posterior is obtained as

\begin{equation}
p_{\text{hybrid}}(c \mid x) = \frac{1}{K} \sum_{k=1}^{K} p_k(c \mid x).
\end{equation}

In the weighted configuration, normalized coefficients $\alpha_k$ satisfying $\sum_{k=1}^{K} \alpha_k = 1$ are introduced:

\begin{equation}
p_{\text{hybrid}}(c \mid x) = \sum_{k=1}^{K} \alpha_k \, p_k(c \mid x).
\end{equation}

The final prediction is determined according to

\begin{equation}
\hat{y} = \arg\max_{c} p_{\text{hybrid}}(c \mid x).
\end{equation}
By aggregating predictions at the probabilistic level, architectural independence is preserved while enabling complementary contributions from heterogeneous backbones.

\paragraph{Hybrid Classification Pipeline}
The inference procedure can be summarized as follows:

\begin{enumerate}
    \item A preprocessed TEM image is supplied as input.
    \item The image is processed independently by each CNN and transformer backbone.
    \item Logits are transformed into posterior probabilities via softmax.
    \item Probability distributions are aggregated using equal or performance-based weighting.
    \item The final class label is selected through maximum posterior estimation.
\end{enumerate}
This formulation allows modular extension to additional models without structural redesign, thereby maintaining computational flexibility.

\subsection{Implementation Details}
All experiments were implemented in PyTorch. Training was conducted on NVIDIA GPU hardware to accelerate computation. Hyperparameters were selected empirically, and random seeds were fixed to ensure reproducibility. Evaluation results and visualizations were recorded for comparative analysis. Model performance was assessed using accuracy, macro-averaged precision, recall, and F1 score. ROC and precision–recall analyses were conducted under a one-versus-rest scheme. Confusion matrices were also examined to analyze inter-class misclassification patterns.

\section{Experimental Results}

\subsection{Performance Comparison}

\begin{table*}[t]
\centering
\caption{Performance comparison of individual and hybrid models on the TEM virus dataset. Metrics are macro-averaged.}
\label{tab:classification_results}
\begin{tabular}{lcccc}
\hline
\textbf{Model} & \textbf{Accuracy} & \textbf{Precision} & \textbf{Recall} & \textbf{F1 Score} \\
\hline
ResNet50 & 0.5887 & 0.4651 & 0.5050 & 0.4501 \\
EfficientNetV2 & 0.7823 & 0.7608 & 0.7478 & 0.7428 \\
ConvNeXt & 0.7581 & 0.7131 & 0.7128 & 0.7066 \\
ViT & 0.8065 & 0.7700 & 0.7685 & 0.7601 \\
DeiT & 0.8669 & 0.8419 & 0.8124 & 0.8135 \\
Swin Transformer & 0.8831 & 0.8475 & 0.8575 & 0.8444 \\
\hline
Hybrid (Equal) & 0.8750 & 0.8534 & 0.8480 & 0.8477 \\
Hybrid (Weighted) & \textbf{0.8831} & \textbf{0.8577} & \textbf{0.8525} & \textbf{0.8528} \\
\hline
\end{tabular}
\end{table*}
The comparative performance is presented in Table~\ref{tab:classification_results}, where six individual models and two hybrid configurations are evaluated. The highest macro-F1 score among standalone architectures was achieved by the Swin Transformer (0.8444), with DeiT demonstrating similar performance. In contrast, convolutional models exhibited reduced balanced performance, most notably in the case of ResNet50, where the impact of class imbalance was evident. With the introduction of hybrid aggregation, improvements in macro-averaged metrics were observed. The performance-weighted configuration achieved the highest macro-F1 score (0.8528) while maintaining strong accuracy (0.8831), indicating improved class-wise balance without compromising overall predictive capability.

\subsection{Confusion Matrix Analysis}
\begin{figure*}[t]
\centering
\includegraphics[width=0.25\linewidth]{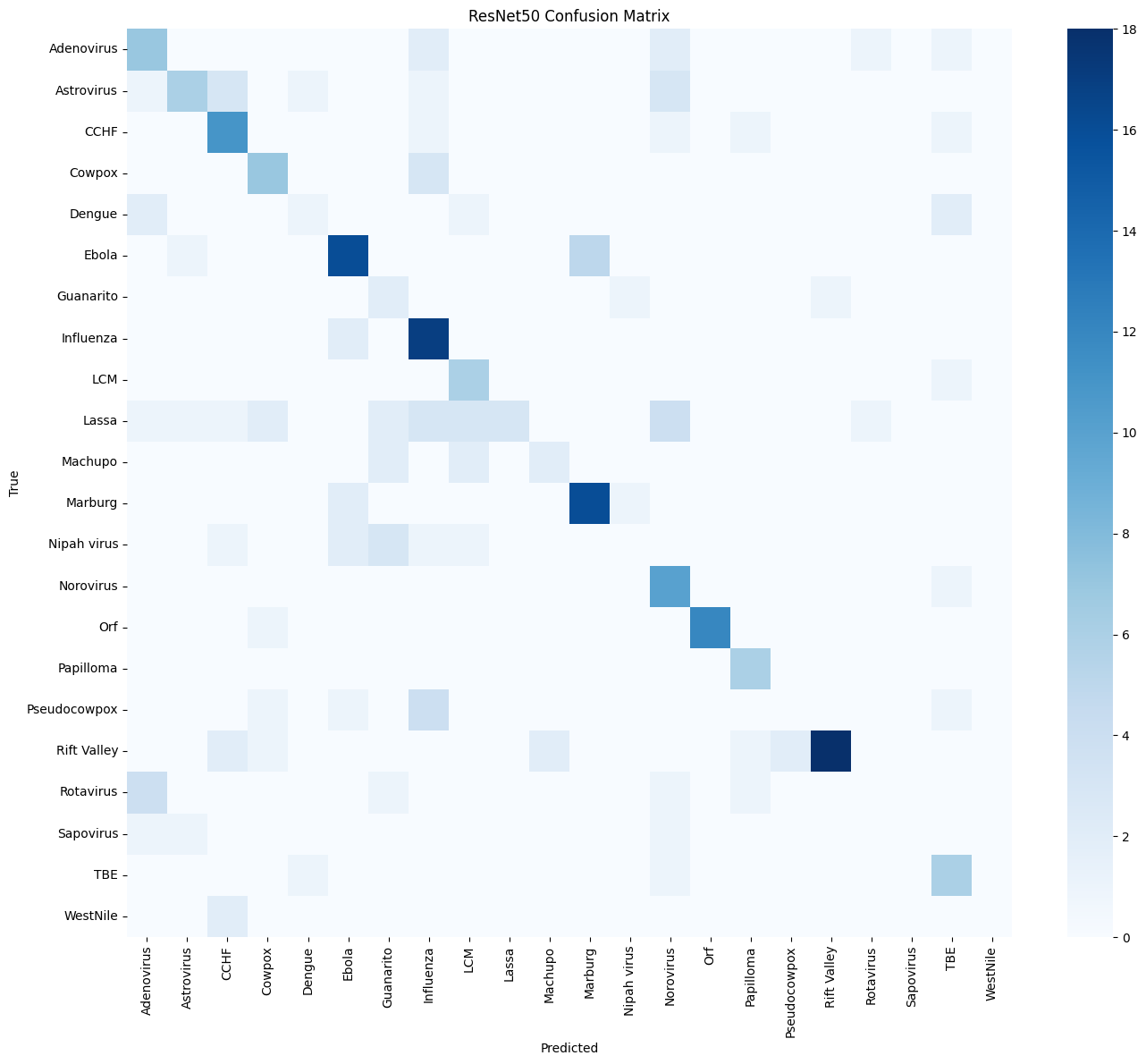}
\includegraphics[width=0.25\linewidth]{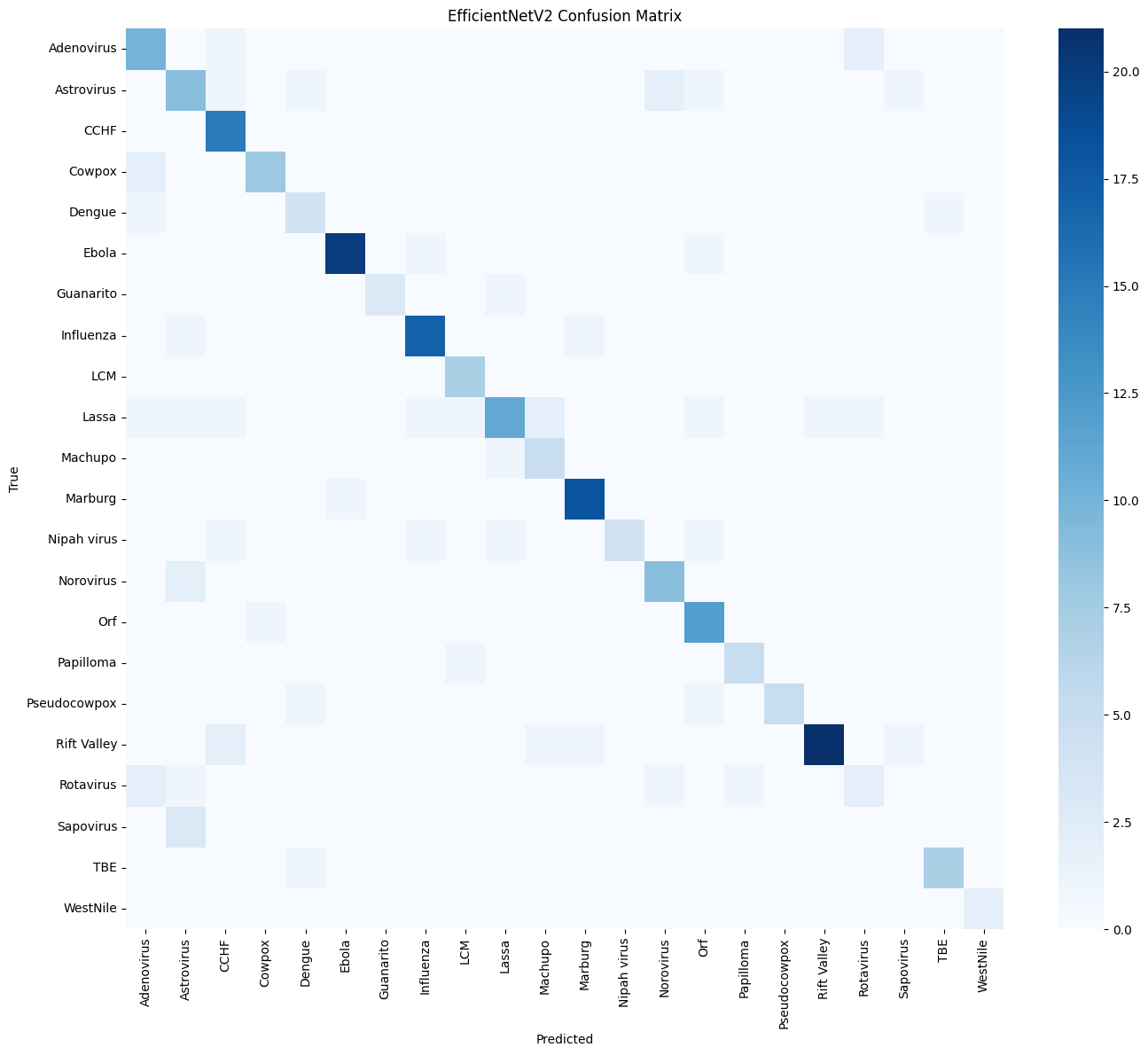}
\includegraphics[width=0.25\linewidth]{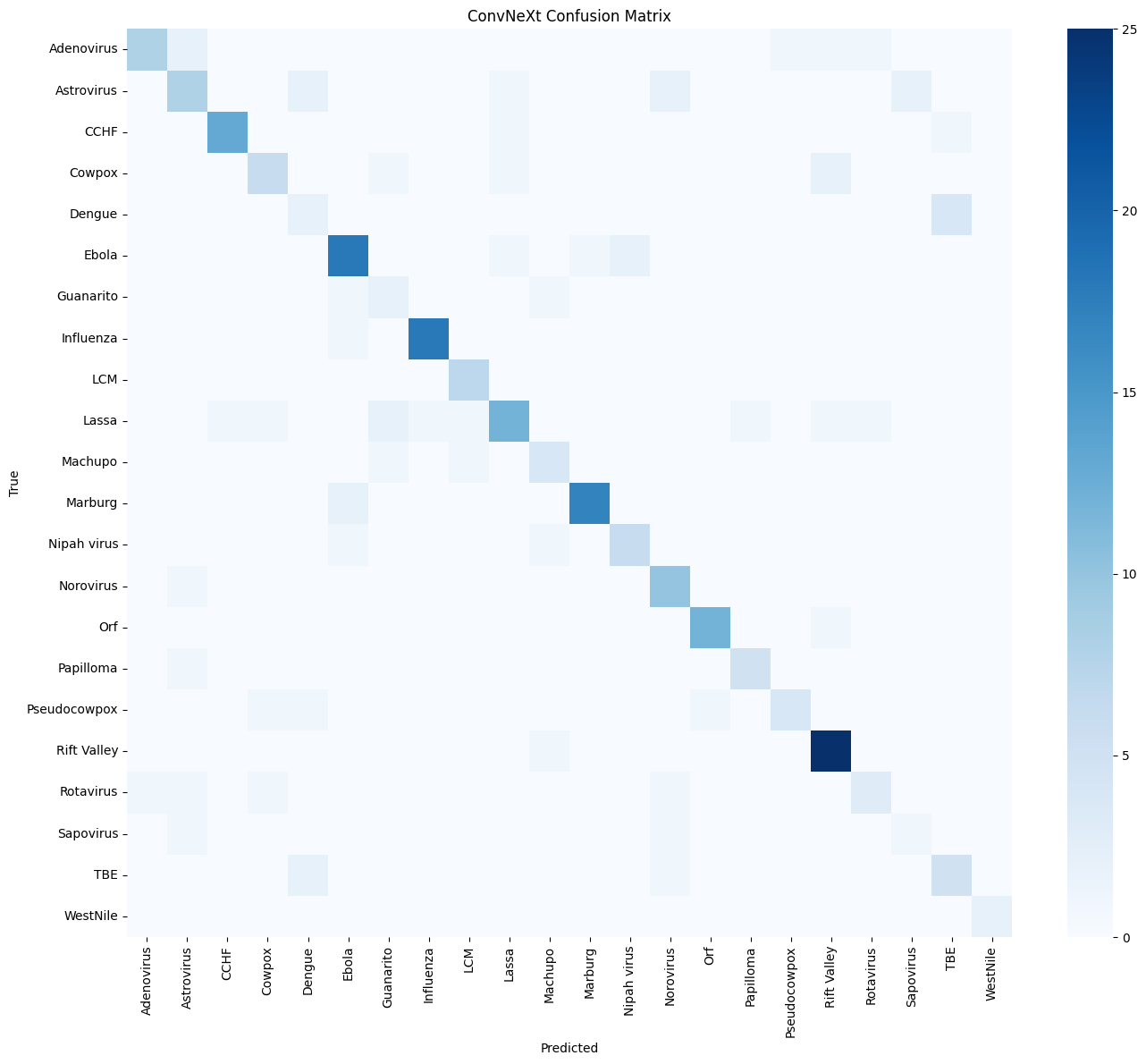}
\caption{Normalized confusion matrices for convolutional architectures: (left) ResNet50, (center) EfficientNetV2, (right) ConvNeXt.}
\label{fig:cm_cnn_group}
\end{figure*}

\begin{figure*}[t]
\centering
\includegraphics[width=0.25\linewidth]{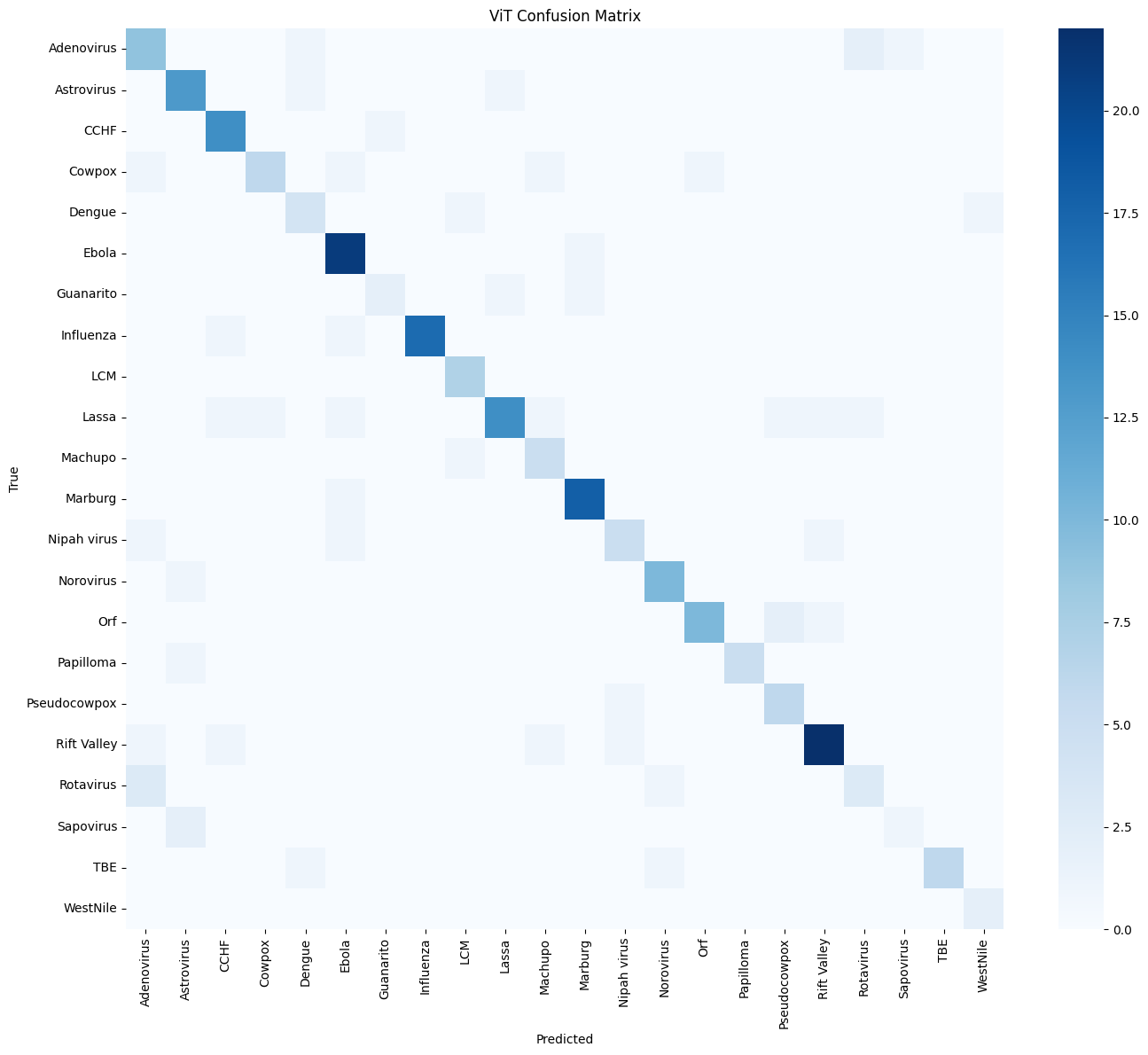}
\includegraphics[width=0.25\linewidth]{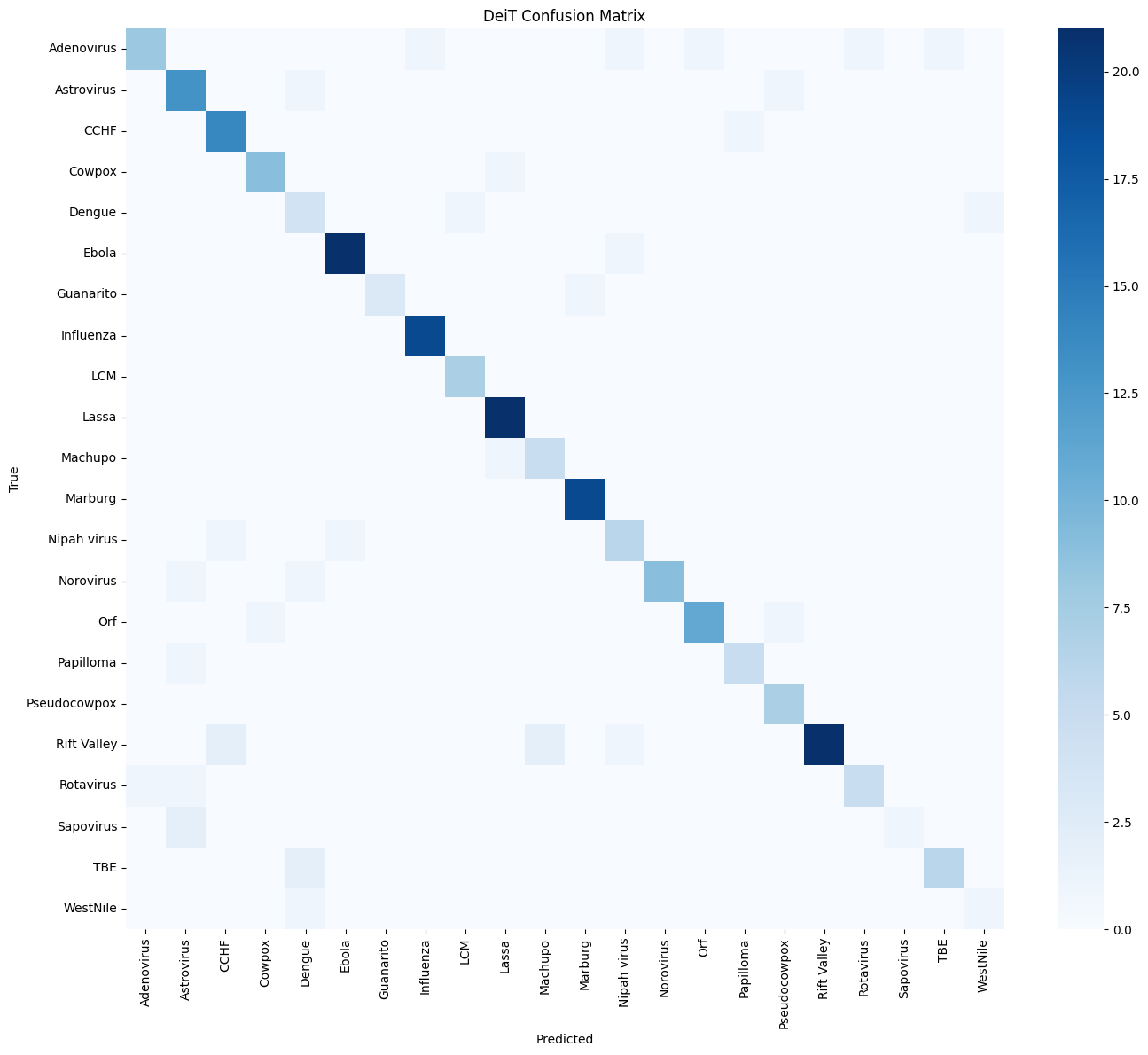}
\includegraphics[width=0.25\linewidth]{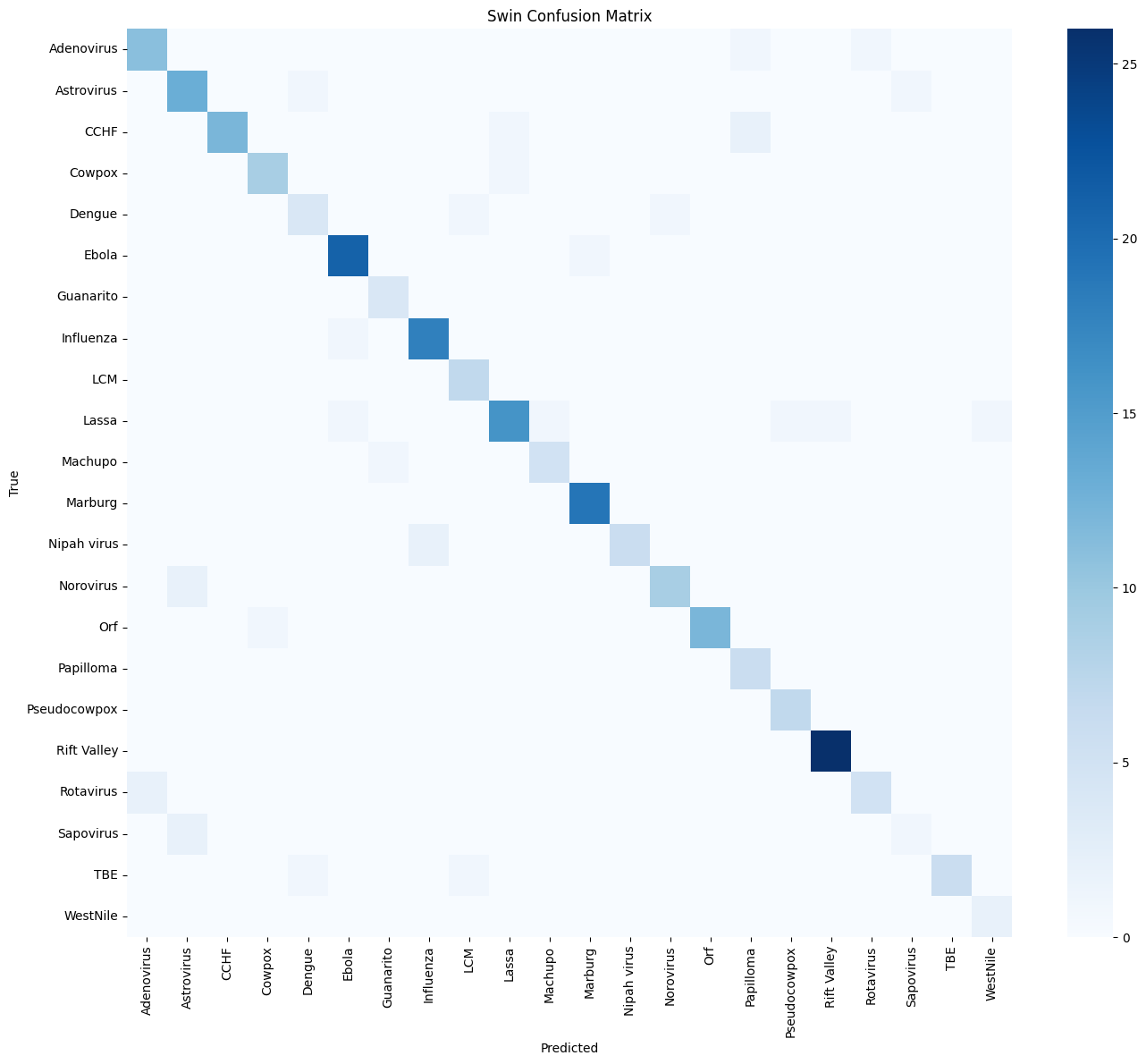}
\caption{Normalized confusion matrices for transformer-based architectures: (left) ViT, (center) DeiT, (right) Swin Transformer.}
\label{fig:cm_transformer_group}
\end{figure*}

\begin{figure*}[t]
\centering
\includegraphics[width=0.30\linewidth]{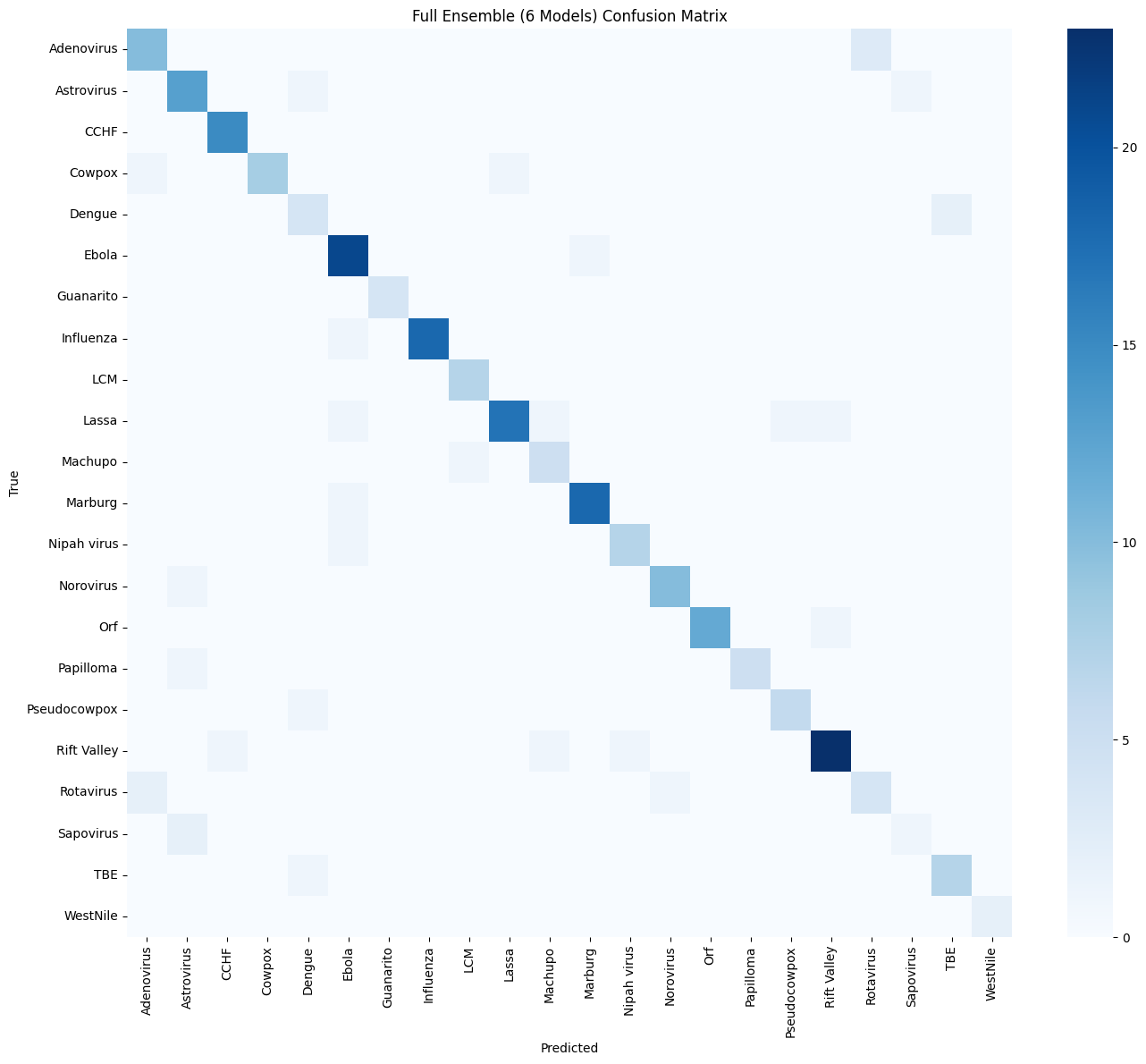}
\includegraphics[width=0.30\linewidth]{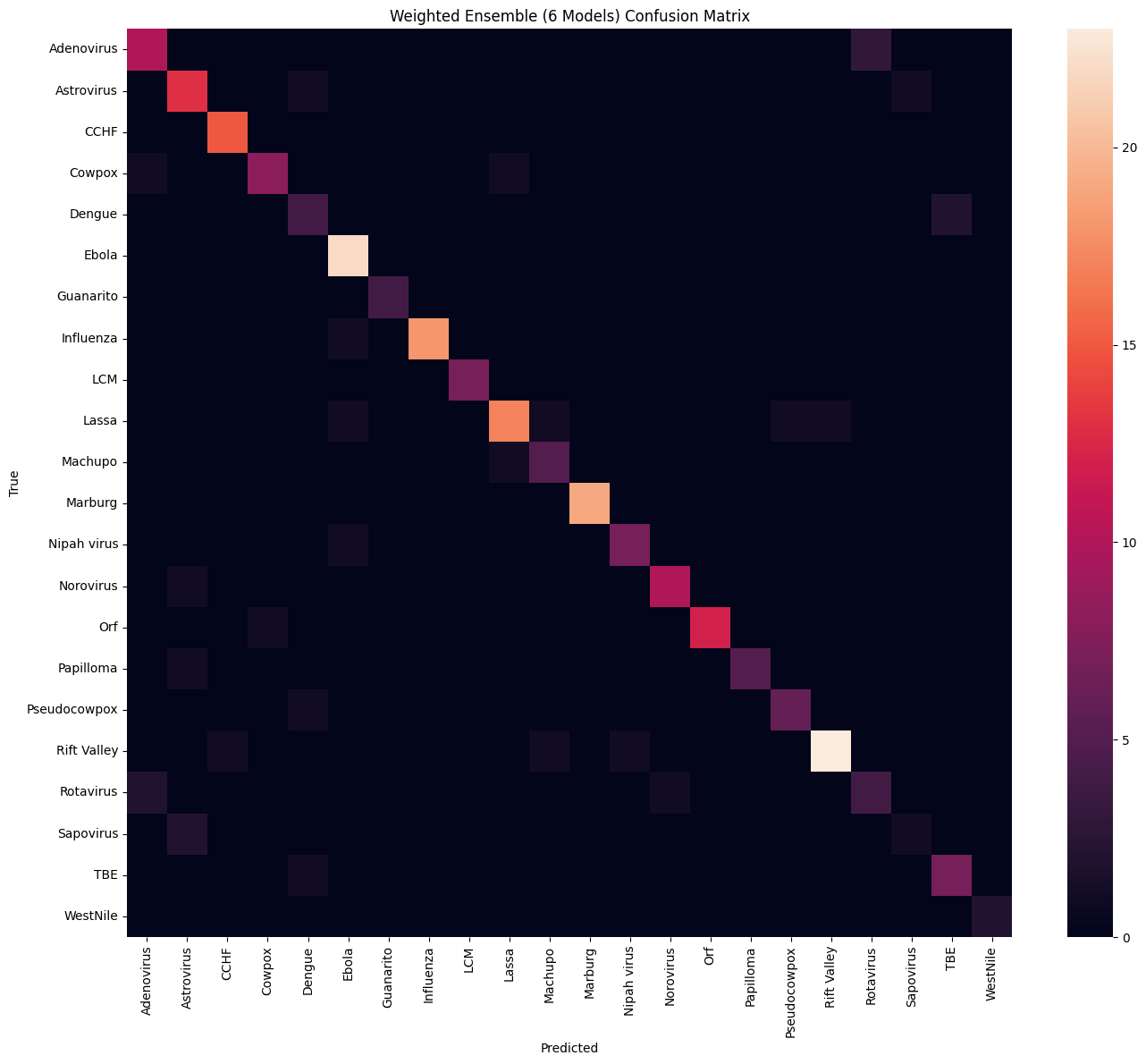}
\caption{Normalized confusion matrices for hybrid configurations: (left) Equal-weight hybrid, (right) Performance-weighted hybrid.}
\label{fig:cm_hybrid_group}
\end{figure*}

Normalized confusion matrices were computed to analyze inter-class confusion patterns, as illustrated in Figures~\ref{fig:cm_cnn_group}–\ref{fig:cm_hybrid_group}. From Figure~\ref{fig:cm_cnn_group}, it may be observed that convolutional architectures exhibit a more dispersed off-diagonal distribution, particularly among morphologically similar virus categories. This phenomenon indicates difficulty in distinguishing subtle structural variations when relying primarily on localized receptive fields. In contrast, the transformer-based models presented in Figure~\ref{fig:cm_transformer_group} demonstrate comparatively stronger diagonal concentration, reflecting improved global contextual modeling. Notably, the Swin Transformer exhibits more consistent per-class recognition stability across minority categories.The hybrid configurations shown in Figure~\ref{fig:cm_hybrid_group} reveal further consolidation of diagonal dominance. In particular, the weighted hybrid configuration reduces confusion among underrepresented classes, suggesting that performance-based probabilistic scaling mitigates individual model bias.

\subsection{Comparative Interpretation}

Considering confusion matrix patterns alongside quantitative metrics, improved classification robustness can be observed with increasing architectural diversity. Transformer-based models were generally found to achieve better balanced performance than conventional CNNs, while further improvements were obtained through decision-level aggregation. The highest macro-F1 score achieved by the weighted hybrid configuration suggests that complementary representations were effectively combined without requiring feature-level fusion. It may therefore be inferred that probabilistic hybrid strategies provide an efficient approach for enhancing class-balanced performance in heterogeneous TEM datasets.

\section{Conclusion}
A comparative evaluation of convolutional and transformer-based architectures was conducted for multi-class virus recognition in transmission electron microscopy imagery. Under consistent training conditions, transformer-based models were observed to achieve stronger macro-averaged performance, particularly in classes with subtle structural variation, whereas convolutional networks showed greater sensitivity to class imbalance. To exploit architectural complementarity without feature-level fusion, decision-level aggregation was applied using equal and performance-weighted averaging. The weighted configuration yielded consistent improvements in macro-F1 score and reduced misclassification among minority classes, as reflected in confusion matrix patterns. Transformer-based representations were found to be advantageous in heterogeneous TEM data, while hybrid aggregation contributed to improved class-wise balance, thereby establishing a comparative baseline for virus classification in electron microscopy imagery.

\bibliographystyle{splncs04}
\bibliography{mybibliography}
\end{document}